\documentclass[journal=jacsat,manuscript=article]{achemso}

\usepackage[version=3]{mhchem} % Formula subscripts using \ce{}
\usepackage{graphicx}
\usepackage{dcolumn}
\usepackage{bm}
\usepackage{comment}
\usepackage{color}

\author{Bianca Turini}
\affiliation{NEST, Istituto Nanoscienze-CNR and Scuola Normale Superiore, 56127 Pisa, Italy}

\author{Sedighe Salimian}
\affiliation{NEST, Istituto Nanoscienze-CNR and Scuola Normale Superiore, 56127 Pisa, Italy}

\author{Matteo Carrega}
\affiliation{CNR-SPIN, 16146 Genova, Italy}

\author{Andrea Iorio}
\affiliation{NEST, Istituto Nanoscienze-CNR and Scuola Normale Superiore, 56127 Pisa, Italy}

\author{Elia Strambini}
\affiliation{NEST, Istituto Nanoscienze-CNR and Scuola Normale Superiore, 56127 Pisa, Italy}

\author{Francesco Giazotto}
\affiliation{NEST, Istituto Nanoscienze-CNR and Scuola Normale Superiore, 56127 Pisa, Italy}

\author{Valentina Zannier}
\affiliation{NEST, Istituto Nanoscienze-CNR and Scuola Normale Superiore, 56127 Pisa, Italy}

\author{Lucia Sorba}
\affiliation{NEST, Istituto Nanoscienze-CNR and Scuola Normale Superiore, 56127 Pisa, Italy}

\author{Stefan Heun}
\email{stefan.heun@nano.cnr.it}
\affiliation{NEST, Istituto Nanoscienze-CNR and Scuola Normale Superiore, 56127 Pisa, Italy}

\title{Josephson Diode Effect in High Mobility InSb Nanoflags}

\begin{document}

\begin{abstract}
 We report evidence of non--reciprocal dissipation--less transport in single ballistic InSb nanoflag Josephson junctions, owing to a strong spin--orbit coupling. Applying an in--plane magnetic field, we observe an inequality in supercurrent for the two opposite current propagation directions. This demonstrates that these devices can work as Josephson diodes, with dissipation--less current flowing in only one direction. For small fields, the supercurrent asymmetry increases linearly with the external field, then it saturates as the Zeeman energy becomes relevant, before it finally decreases to zero at higher fields. We show that the effect is maximum when the in--plane field is perpendicular to the current vector, which identifies Rashba spin--orbit coupling as the main symmetry--breaking mechanism. While a variation in carrier concentration in these high--quality InSb nanoflags does not significantly influence the diode effect, it is instead strongly suppressed by an increase in temperature. Our experimental findings are consistent with a model for ballistic short junctions and show that the diode effect is intrinsic to this material. Our results establish InSb Josephson diodes as a useful element in superconducting electronics.
\end{abstract}

%%%%%%%%%%%%%%%%%%%%%%%%%%%%%%%%%%%%%%%%%%%%%%%%%%%%%%%%%%%%%%%%%%%%%
%% Start the main part of the manuscript here.
%%%%%%%%%%%%%%%%%%%%%%%%%%%%%%%%%%%%%%%%%%%%%%%%%%%%%%%%%%%%%%%%%%%%%

\section{Introduction}

Non--reciprocal charge transport is at the heart of conventional electronics, in which a fundamental building block, the diode, is based on the $p-n$ junction. In such systems, the rectification effect takes place due to the presence of a heterojunction that explicitly breaks inversion  symmetry. Only very recently it has been proposed that the superconducting analogue of non--reciprocal transport can be made \cite{Hu2007}, based on similar symmetry arguments: in this case, rectification is expected when time-reversal and inversion symmetries are simultaneously broken. This scenario has been investigated in condensed matter setups, both for fundamental reasons and technological applications \cite{Wakatsuki2017,Qin2017,Tokura2018,Hoshino2018,Yasuda2019}. It is worth to note that supercurrent rectification has been achieved before in superconducting quantum interference devices (SQUIDs), where the flux tunability allows to reach high rectification coefficients useful for applications. However, this rectification is of extrinsic nature and not an intrinsic property, being induced by asymmetric junctions and the presence of an external flux threading the SQUID \cite{Barone1982,Paolucci2019,Souto2022}.

In fact, an intrinsic supercurrent analogue exists --- the supercurrent diode effect (SDE) --- whose exploitation would constitute a real breakthrough for low temperature technology and superconducting electronics. The first experimental report on the SDE, based on electrically polar materials \cite{Ando2020}, has appeared very recently, demonstrating supercurrent rectification. Soon after, other systems \cite{Miyasaka2021,Baumgartner2022,Wu2022,Baumgartner2022a,Strambini2022,DiezMerida2021,Bauriedl2021,Shin2021,Lin2021,Pal2021,Hou2022,Gupta2022} have been inspected looking at supercurrent non--reciprocal transport, complemented by theoretical efforts \cite{Kopasov2021,Misaki2021,Daido2022,Halterman2022,Scammell2022,Yuan2022,He2022,Ilic2022,Zhang2021,Davydova2022,Souto2022}, in order to shed light on the microscopic mechanisms responsible for the SDE.  However, both from an experimental and a theoretical point of view, this field is still in its infancy with several open questions.

During the last decade, there has been a widespread interest in the physics of hybrid systems comprising superconductors and low-dimensional semiconductors featuring strong spin-orbit coupling (SOC).
Indeed, these systems offer an ideal platform to develop new architectures able to coherently control electron spin with great impact in spintronics and topological quantum computing \cite{Qu2016,Chen2021,Baumgartner2022a,Fuchs2022}.

Exploiting the large SOC of InAs, the authors of Ref.~\citenum{Baumgartner2022} have observed SDE in an array of Josephson junction (JJ) devices. There, SDE has been observed by measuring the variation of the circuit kinetic inductance upon the application of an in-plane magnetic field. However, a direct measurement of the non--reciprocal supercurrent transport in a single JJ based on a strong SOC semiconductor, and its tunability by external means, is still lacking. The observation of SDE in planar JJs has also been reported in hybrid systems comprising Dirac materials like twisted bilayer and trilayer graphene or topological semimetals \cite{DiezMerida2021,Lin2021,Pal2021}.

Supercurrent rectification in hybrid Josephson junctions has been also referred to as Josephson diode effect (JDE). Here, the combination between SOC and superconducting proximity leads to a strong interaction between spin, charge, and superconducting phase, which is the working principle of the $\varphi_0$ junction. In such devices, the current-phase relation (CPR) is shifted by an anomalous phase $\varphi_0$ which is externally controllable \cite{Strambini2020}. Moreover, $\varphi_0$ junctions can be considered the precursors of the Josephson diode: as shown in Ref.~\citenum{Baumgartner2022}, highly transmissive junctions, which operate in the short-junction regime, are characterized by a skewed current-phase-relation, which leads to supercurrent rectification in the presence of an anomalous phase shift.

In this context, Indium Antimonide (InSb) represents a valid platform. InSb has a narrow band gap (0.23 eV) \cite{Qu2016,Moehle2021,Chen2021a}, a small effective mass ($m^* = 0.018 \; m_e$) \cite{Sladek1957,Vurgaftman2001,Qu2016,Mata2016,Ke2019,Vries2019,Lei2021}, and exhibits a strong SOC and a large Land\'e g-factor ($\left| g^* \right| \sim 50$)\cite{Litvienko2008}. In InSb 2D nanostructures, a similar value is measured in the out-of-plane direction, while the in-plane value $g^*_{ip}$ is reduced by a factor 2, independently on the crystallographic direction ($g^*_x \sim g^*_y \sim 25$) \cite{Qu2016,Mata2016,Lei2021}. Moreover, a Rashba spin--orbit strength of $\alpha_R \sim 0.42$~eV \AA{} was reported for InSb nanosheets \cite{Chen2021}, which yields a spin--orbit energy of $E_{so} = \left( m^* \alpha_R^2 \right) / \left( 2 \hbar^2 \right) \sim 200 \; \mu$eV.

In this work, relying on low--temperature magneto--transport measurements, we present the first report of JDE in single planar JJs based on high-quality InSb nanoflags \cite{Verma2020,Verma2021}. These structures have been used to form ballistic planar JJs, upon deposition of superconducting contacts \cite{Vries2019,Salimian2021}. Owing to their intrinsic strong SOC and sizable superconducting proximity \cite{Mata2016,Pan2016,Gazibegovic2019,Verma2020,Verma2021}, they become a natural platform to investigate JDE and to obtain insight on its microscopic mechanism.

Previous experiments on analogous devices show a Nb-induced gap of $\Delta^*= 160$ $\mu$eV \cite{Salimian2021}. This value is close to the predicted SOC energy $E_{so}$, suggesting that SOC plays a relevant role in the physics of these InSb JJs. The high quality of the material is a crucial feature that permits to work in the ballistic regime, allowing for the direct observation of a non-reciprocal supercurrent. In addition, the dependence of the JDE with respect to external parameters can provide important information on the symmetry--breaking mechanisms at play. Our observations are consistent with a dominant Rashba coupling related to structural inversion asymmetry. We provide a direct demonstration of JDE in a single and scalable planar JJ, which constitutes an important step forward in the understanding of the JDE mechanism and in the pursuit for new and low-power electronic devices based on superconducting circuits. 

\section{Results and discussion}

\subsection{\label{sec:Samgeo}Sample characterization}

The system under investigation is a superconducting--normal metal--superconducting (SNS) planar Josephson junction, where the N region consists of an InSb quasi-2D nanostructure. Previous characterization showed that these InSb nanoflags are defect--free zinc--blende structures with high mobility (up to $\mu_e \sim 29 500$~cm$^2$V$^{-1}$s$^{-1}$) and a large mean free path ($\lambda_{e} = 500$~nm) at $T=4.2$~K \cite{Verma2021}. We notice that the extracted Fermi wavelength ($\lambda_F = \sqrt{2 \pi / n} \sim 30$~nm for carrier concentration $n = 8.5 \times 10^{11}$~cm$^{-2}$)\cite{Verma2021} is of the same order of magnitude as the thickness of the nanoflags ($t\approx 100$~nm)\cite{Verma2021,Salimian2021}: by evaluating the number of the active transport modes ($\approx 40$)\cite{Salimian2021} and the degeneracy of the vertical subbands, we find that these devices are well placed in the quantum limit with a clear 2D character.

\begin{figure}[thp]
    \includegraphics[width=0.5\columnwidth]{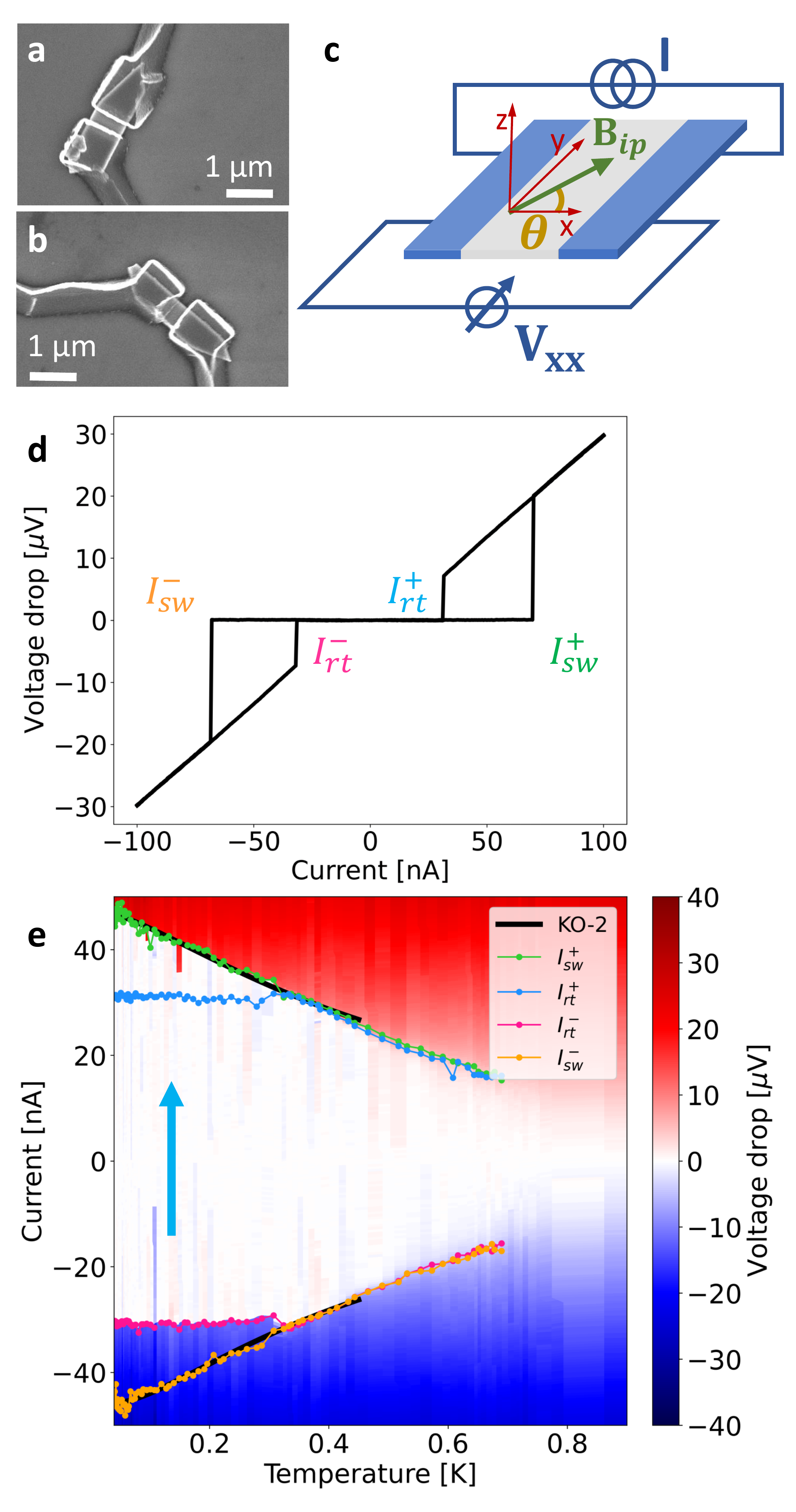}
    \caption{\label{fig:figure1} Device characterization. (a,b) SEM images of the two devices, G4 and G5. (c) Sketch of the measurement schematics. Also the angle $\theta$ between the orientation of the in-plane magnetic field $B_{ip}$ and the direction of current flow $I$ is indicated. (d) $V-I$ characteristics at $T=30$~mK. The difference between switching and retrapping current, defined in the plot, is clearly visible. (e) Temperature dependence of the $V-I$ characteristics. The individual spectra were measured sweeping from negative to positive bias values, in the direction indicated by the arrow. Thus, the 2D plot shows the switching trace for positive values of the current and the retrapping trace for the negative ones. The extracted values of $I^+_{sw},I^+_{rt},I^-_{sw},I^-_{rt}$ for each temperature and both sweep directions are shown in green, blue, pink and orange, respectively. The black lines show a fit of the switching currents to the KO-2 model (see text). For (d,e): device G4, $B=0$~mT, and $V_{bg}=40$~V.}
\end{figure}

For device fabrication, the nanoflags are placed on a $p$--doped Si/SiO$_2$ substrate, which acts as a global back--gate. The nanoflags are contacted by 10/150~nm of Ti/Nb, which defines the superconducting leads, leaving the central region of the nanoflag uncovered. The dimensions of the resulting planar Josephson junctions, length $L=200$~nm and width $W=700$~nm, are such that the devices work in the ballistic regime. More details on device fabrication can be found in the supplemental material of Ref.~\citenum{Salimian2021}. Figure~\ref{fig:figure1}a and Figure~\ref{fig:figure1}b show the two devices discussed in this manuscript, named G4 and G5, respectively. The two devices, resulting from the same fabrication process, are characterized by the same material and geometric parameters. The superconducting coherence length can be determined as $\xi_s= \hbar v_F/\Delta$ \cite{Lee2018,Vries2019,Zhi2019,Banszerus2021}, where $\Delta$ is the gap in the superconductor and $v_F$ the Fermi velocity in the semiconductor. By inserting the value of the Nb gap and the value of the Fermi velocity of the InSb nanoflags, we obtain a coherence length much larger than the length of the uncovered region ($\xi_s \approx 750$~nm $> L$). Thus, the devices operate in the short junction regime.

Transport measurements were performed in an Oxford Triton 200 dilution refrigerator with a base temperature of 30~mK. The measurement setup is sketched in Figure~\ref{fig:figure1}c. We study the low-temperature magneto-transport of the devices in the presence of an in-plane magnetic field. A relative angle $\theta$ $\left( \pm 180^\circ \right)$ can be set between the orientation of the in-plane magnetic field and the direction of current flow, $\vec{B}_{ip}$ and $\vec{I}$, respectively. The sign of $\theta$ is given by the direction of the $\vec{B}_{ip} \times \vec{I}$ vector. With this definition, $\theta = 0^\circ$ for $\vec{B}_{ip} \parallel \vec{I}$ and $\theta = 90^\circ$ for $\vec{B}_{ip} \perp \vec{I}$.

Figure~\ref{fig:figure1}d shows a characteristic $V-I$ curve of device G4 measured at $T=30$ mK and $V_{bg}=40$ V. We can clearly distinguish the switching ($I_{sw}$) and the retrapping ($I_{rt}$) currents.\cite{footnote1}
%\footnote{We note that the data of Ref.~\citenum{Salimian2021}, for which this hysteretic behavior was not observed, were obtained at $T = 250$~mK, i.e., at a higher temperature than employed here.}
Figure~\ref{fig:figure1}e shows that the extent of the superconducting region decreases monotonically with increasing temperature. The data in the 2D plot are collected by performing a sweep from negative to positive bias, so that the upper plane shows the switching current, while the lower one presents the retrapping value. We also measured the opposite sweep direction (from positive to negative bias, shown in Section I of the Supporting Information). For both sweep directions, we extracted the values of $I_{sw}$ and $I_{rt}$. They are shown as dots in Figure~\ref{fig:figure1}e, overlaid on the 2D plot. The values of $I_{sw}$ and $I_{rt}$ differ for temperatures lower than $\sim 300$~mK, consistent with previous measurements \cite{Salimian2021}. This hysteretic behavior is typical of SNS weak-links \cite{Tinkham1996,Guiducci2019a} and is commonly understood as Joule heating of the N region in the dissipative regime \cite{Schapers2001,Courtois2008,Fornieri2013a,Guiducci2019}. In all the following arguments, the switching and retrapping currents are considered separately.

As shown in Figure~\ref{fig:figure1}e, the temperature dependence of the switching current is well described by the Kulik--Omelyanchuk model in the clean limit (KO--2) \cite{Kulik1977,Golubov2004,Lee2015}. This confirms that the devices are in the ballistic short--junction regime, which leads to a skewed CPR including higher harmonics, crucial for observing the JDE \cite{Baumgartner2022}. The extracted value for the induced gap is $\Delta^*= 108 \pm 4 \; \mu$eV, consistent with the values found in literature \cite{Salimian2021}. The resulting transmission probability $\tau\sim 0.99$ confirms the high quality of the interfaces in the devices, as previously reported \cite{Salimian2021}. More details on the model are given in Section II of the SI.

\subsection{\label{sec:SDE}Josephson Diode Effect}

\begin{figure}[hpt]
    \includegraphics[width=0.7\columnwidth]{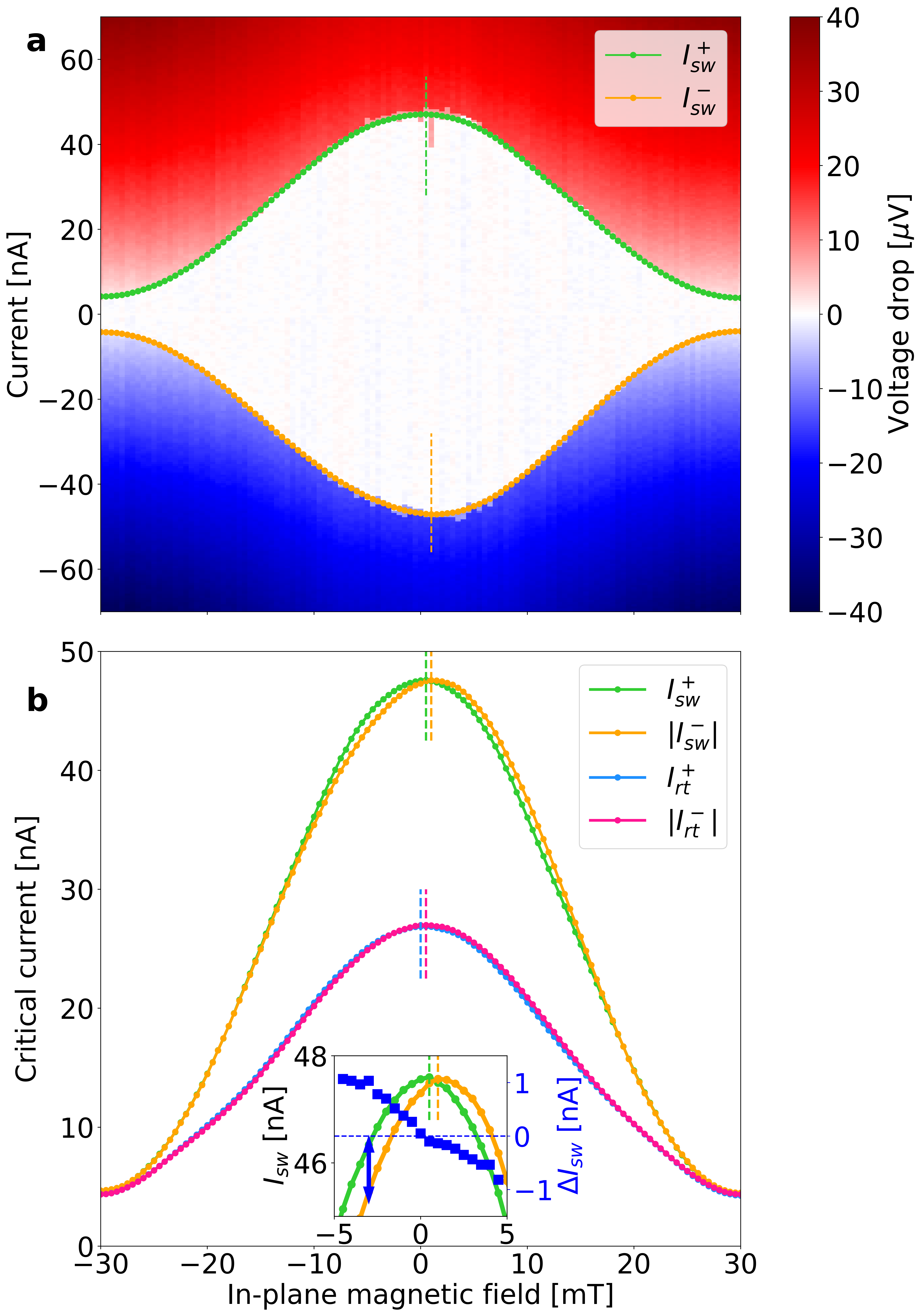}
    \caption{\label{fig:figure2} Switching current dependence on in-plane magnetic field. (a) Voltage drop across the junction versus applied current bias $I$ and in-plane magnetic field $B_{ip}$. The green (orange) dots indicate the positive (negative) switching currents, as defined in the main text. (b) The switching current demonstrates a clear asymmetry between the positive and negative branches, shown in green and orange, respectively. The inset shows a zoom--in to the region around $B_{ip} \sim 0$, to better visualize the differences in switching current. It also shows that $\Delta I_{sw} = I_{sw}^+ - \left| I_{sw}^- \right|$ changes linearly with in--plane magnetic field around $B_{ip} = 0$. The blue and pink lines correspond to the positive and negative retrapping current. In both cases, the maximum of the curves is different: the negative branch is higher for positive values of the magnetic field, and the relation is reversed for negative field. Device G5, angle $\theta=129^{\circ}$, $T=30$~mK, and $V_{bg}=40$~V.}
\end{figure}

We now report on the evidence of the Josephson diode effect in our devices, showing that it only requires an in--plane magnetic field orthogonal to the direction of current flow. We chose to work at $T=30$ mK and $V_{bg}=40$~V, where the nanostructures are highly conducting, to have the maximum switching current (see Ref.~\citenum{Salimian2021} and Figure S7 of the SI). Figure~\ref{fig:figure2}a shows the voltage drop across the junction G5 versus applied current bias $I$ and in-plane magnetic field $B_{ip}$, with the relative angle set to $\theta=129^{\circ}$ (cf.~Figure~\ref{fig:figure1}c). The data was taken by increasing the bias from zero to positive (negative) values, in order to exclude a current heating of the device before the switching event. The superconducting region, defined by dissipation--less charge transport, corresponds to the white area. It can be seen that the supercurrent is maximum around zero in--plane magnetic field and decreases with increasing field until $B_{ip} = \pm 30$~mT, for which it is nearly but not completely suppressed.

From the data, positive and negative switching currents can be extracted. The values of positive switching current $I_{sw}^+$ and negative switching current $I_{sw}^-$ are included in Figure~\ref{fig:figure2}a as green and orange dots, respectively. Careful analysis of this data shows that the pattern is slightly skewed with opposite polarity for the two sweep directions. The position of the maximum (minimum) value of $I_{sw}^+$ ($I_{sw}^-$) is indicated in the panel by dashed lines. Note that the two sweep directions are measured consequentially for each value of $B_{ip}$, hence a simple residual magnetization could not explain the opposite skewness of the two patterns. Interestingly, the maximum of the switching current is not observed for zero magnetic field, as one would expect for a standard Fraunhofer-like pattern, but is slightly shifted to a finite magnetic field whose sign depends on the sweeping direction. Therefore we observe that the magnitude of the positive (negative) switching current \textit{increases} with respect to the value at zero field for small negative (positive) values of the magnetic field.

The asymmetry between the positive and negative branches is more clearly visible by comparing the absolute values of the two curves, shown in Figure~\ref{fig:figure2}b. For negative magnetic field, $I_{sw}^+ \ge \left| I_{sw}^- \right|$, while for positive magnetic field, $I_{sw}^+ \le \left| I_{sw}^- \right|$. Thus, for non-zero values of $B_{ip}$, there exists a range of bias current values for which the transport across the junction is non-dissipative only in one direction, indicating the presence of JDE. In addition, the action of the Josephson rectifier is reversed with the sign of the magnetic field. In the same measurement, also the retrapping current was recorded when sweeping the current back to zero after each switching event. These data are shown in Figure~\ref{fig:figure2}b, as well. The same JDE can be observed, albeit with smaller magnitude, between the two branches of the retrapping current. Qualitatively identical results were also observed for device G4, as reported in Section III of the SI.

\begin{figure*}[hpt]
    \includegraphics[width=\columnwidth]{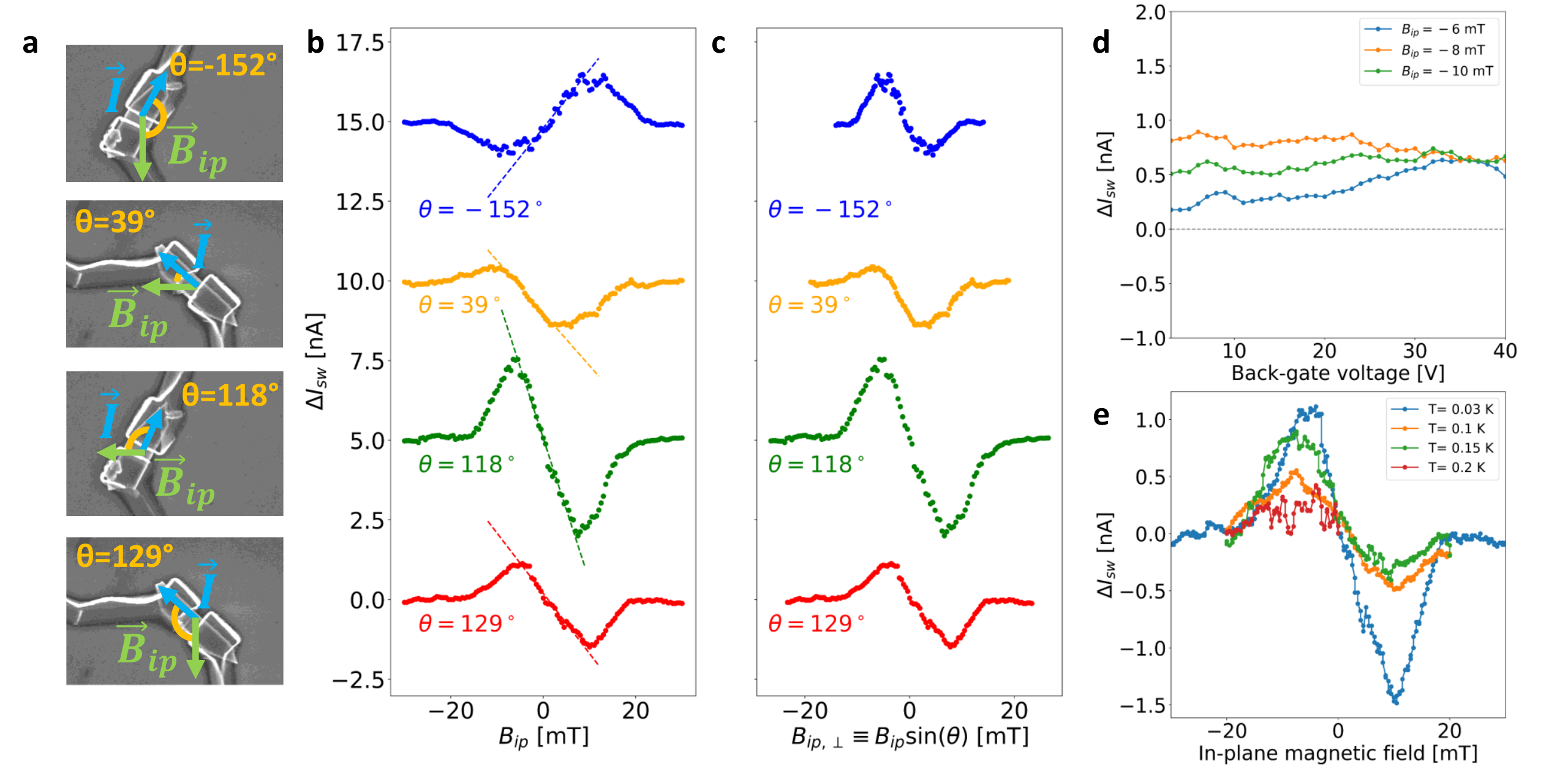}
    \caption{\label{fig:figure4} Behavior of the JDE with in-plane magnetic field, perpendicular component of the in-plane field, back--gate voltage, and temperature. (a) SEM images indicating the relative angle between $B_{ip}$ and $I$ for the measurements shown in (b) and (c). (b) Asymmetry versus in-plane field for different orientations of the devices. Here, the blue and green curves correspond to device G4, while the yellow and red curves correspond to G5. (c) Asymmetry versus the component of the magnetic field perpendicular to the current flow. The maximum of $\Delta I_{sw}$ is observed for $B_{ip, \perp} = -6$~mT for each curve. The amplitude of the effect is maximum when $\theta$ is close to $\pm 90^\circ$, i.e., when the in--plane magnetic field is perpendicular to the current vector, as explained in the main text. Note that the polarity of the curve at $\theta=-152^\circ$ in (c) is reversed with respect to (b) due to the sign of $\sin(\theta)$. In (b) and (c), the curves are shifted by 5~nA each for clarity. (d) Asymmetry versus back--gate voltage, for three different values of the applied in--plane magnetic field. (e) Temperature--dependence of the asymmetry. For (d,e): Device G5.}
\end{figure*}

We use the difference in the switching currents $\Delta I_{sw} = I_{sw}^+ - \left| I_{sw}^- \right|$ to quantify the JDE. The dependence of $\Delta I_{sw}$ on magnetic field is presented in Figure~\ref{fig:figure4}b. The curves are obtained as switching current differences between two consecutive bias sweeps, one in the positive direction, the other in the negative direction. To consider the asymmetry beyond the fluctuations due to stochastic switching, we performed a gentle smoothing procedure, as described in Section IV of the SI. The experiment was repeated for different relative orientations of the magnetic field, as sketched in Figure~\ref{fig:figure4}a, to collect information about the angular dependence of the JDE. The first information to note from Figure~\ref{fig:figure4}b is that all measurements show antisymmetric curves, i.e., $\Delta I_{sw} \left( B_{ip} \right) = - \Delta I_{sw} \left( - B_{ip} \right)$. Furthermore, the curve for $\theta = -152^\circ$ is flipped with respect to the others, so that the sign of the point in magnetic field of maximum value in $\Delta I_{sw}$ is opposite. This is consistent with the different orientations of the devices with respect to the field direction, i.e., the polarity of the $\Delta I_{sw}$ curves reflects the sign of the angle $\theta$, which suggests that $\Delta I_{sw} \propto \vec{B}_{ip} \times \vec{I}$. Secondly, we observe that $\Delta I_{sw}$ varies smoothly from a linear regime around zero field via a smooth rounded maximum at intermediate field values to the high magnetic field region, in which the effect is completely suppressed. The general trend, i.e., the linear behavior near $B_{ip}=0$ and the presence of a maximum, is consistent with previous experimental evidence, in which however a more rapid quenching with increasing field was observed~\cite{Baumgartner2022}. To highlight the linear regime around zero field, we have added linear best fits at the origin of each curve. A similar and consistent behavior was also observed for the retrappig current, as shown in Section V of the SI.

To disentangle the contributions of the parallel ($B_{ip,\parallel}$) and perpendicular ($B_{ip,\perp}$) components of the field, computed with respect to the direction of current flow, we mapped the measured data on the effective $B_{ip,\perp}$. This is shown in Figure~\ref{fig:figure4}c. Note that, in case of $\theta<0$, the change in polarity is due to sign of $\sin(\theta)$. In all data sets, the maximum value of the asymmetry is observed for $B_{ip,\perp} = -6$~mT, while the magnitude of the effect depends on the specific orientation. This shows that the main contribution to the effect is given by the perpendicular component of the field, consistent with the observation that $\Delta I_{sw} \propto \vec{B}_{ip} \times \vec{I}$.

Next, we study the dependence of the JDE on back--gate voltage. By setting the value of the field near the maximal $\Delta I_{sw}$ value of Device G5, we performed back--gate sweeps to the pinch--off of the devices. As shown in Figure~\ref{fig:figure4}d for device G5, the asymmetry is nearly constant in the explored range, which implies that the applied electrical field is not strong enough to significantly modulate spin--orbit coupling, consistently with results in similar systems \cite{Baumgartner2022}. On the other hand, the back gate modulates the carrier concentration very efficiently in these devices \cite{Verma2021}, resulting in a reduction of the switching current from $\sim50$~nA to pinch--off in the same range \cite{Salimian2021}, see also Section VII in the SI. The strength of the effect shown in Figure~\ref{fig:figure4}d is consistent with Figure~\ref{fig:figure4}b. We add that a back--gate sweep performed at zero magnetic field gave a zero asymmetry.

Finally, in Figure~\ref{fig:figure4}e we show the influence of temperature. We performed the same magnetic field sweep at four different temperatures, shown in Section VIII of the SI. The amplitude of the asymmetry is rapidly reduced with increasing temperature and strongly suppressed already for $T=200$~mK. We note that the acquisition at $T=150$ mK is less antisymmetric, which we attribute to stochastic noise. Remarkably, while the diode effect disappears, the switching current at $T=200$~mK is only reduced by about 20\% with respect to its value at base temperature. On the other hand, the magnetic field value at which the maximum value of $\Delta I_{sw}$ is observed, does not depend on temperature.

The same measurement as in Figure~\ref{fig:figure2}a is repeated in an out--of--plane magnetic field (no in--plane component), as shown in Section IX of the SI. In this case, no asymmetry is observed, consistent with previous results for similar systems \cite{Zhi2019,Zhi2019a,Vries2019,Salimian2021}. Finally, we add that all measurements performed at $B = 0$ (for example, Figure~\ref{fig:figure1}d) resulted in asymmetry values equal to zero within the noise level.

\subsection{\label{sec:disc}Discussion}

In the presence of an in-plane magnetic field, these SNS junctions act as superconducting diodes. In Ref.~\citenum{Rasmussen2016}, it has been shown that \textit{either} the presence of an in--plane field parallel to the current direction and a Dresselhaus SOC, \textit{or} an in--plane field perpendicular to the charge flow and a Rashba spin-orbit contribution is a sufficient condition for this effect to emerge. Thus, the determination of which parameter actually drives the JDE provides important information about the key acting mechanisms in the junction. Here, we have measured the JDE for different angles $\theta$, i.e. for different relative strength of the two in-plane components. We have shown in Figure~\ref{fig:figure4}c that the magnitude of the effect increases with the sine of the relative angle, i.e., with the perpendicular component of the in--plane field. Since the effect of this component is mediated by the Rashba coefficient, we can state that here a major role is played by the Rashba SO interaction, consistent with the large effective g--factor of InSb. On the other hand, as shown in Section VI of the SI, we have observed no clear trend with the parallel component, indicating that the Dresselhaus term is of little relevance in this system, consistent with previous results reported for InAs-based JJs \cite{Baumgartner2022a}. We add that we have observed no JDE effect for an applied out--of--plane magnetic field $B_z$, in agreement with these conclusions.

Our experimental evidence presented in Figure~\ref{fig:figure4}c shows that the behavior of $\Delta I_{sw}$ is anti--symmetric with respect to $B_{ip, \perp}$, and its maximum value is reached for $B_{ip, \perp}=-6\pm1$~mT, independent of back--gate voltage, temperature, or the relative angle $\theta$. On the other hand, the magnitude of the effect does depend on the relative angle. The analysis in Figure~\ref{fig:figure4}b shows that the asymmetry depends linearly on the in-plane magnetic field near $B_{ip}=0$, consistent with previous experimental results \cite{Ando2020} and theoretical predictions \cite{Yuan2022}.

To investigate the physics of this system, we consider models for the JDE in short ballistic junctions \cite{Dolcini2015,Davydova2022}. These models are based on the idea of finite momentum Cooper pairs \cite{Hart2017,Ke2019,Yuan2022}, akin to a so-called Doppler shift. In the nanoflag-based system, the magnetic field introduces a Zeeman splitting term and, due to the strong SOC of the material, determines a spatially-varying order parameter in the junction \cite{Hart2017}. Consequently, the Cooper pairs acquire a finite momentum $q$ in the direction perpendicular to the magnetic field and the SO vector. This breaks the equivalence between the two propagation directions $I^+$ and $I^-$ of the current, which, instead, is respected in conventional Josephson systems. We remark that the spatial modulation occurs in the normal region of the junction and not in the superconducting leads.

If the spin--orbit energy $E_{so}$ is much larger than the Zeeman energy $E_z = g_{ip}^* \mu_B B_{ip} \ll E_{so}$, energy bands of opposite spin are split, and a finite Cooper pair momentum is expected \cite{Hart2017}. This condition is fulfilled here, since $E_z = 15$~$\mu$eV at 10~mT and thus much smaller than $E_{so} = 200$~$\mu$eV. In this case, $q v_F = E_z$, with $v_F$ the Fermi velocity, and thus $q \propto B$ \cite{Dolcini2015,Hart2017}. The difference between the magnitudes of the critical currents in opposite directions $\Delta I_{c} = I_{c}^+ - \left| I_{c}^- \right|$ can then be calculated for small $B$ and zero temperature as \cite{Davydova2022}
\begin{equation}
   \Delta I_{c} =  \frac{4eqv_F}{\pi \hbar} + O(B^2).
\end{equation}
In a similar way, up to first order in the magnetic field, we obtain
\begin{equation}
   I_{c}^+ + \left| I_{c}^- \right| =  \frac{2 e \Delta^*}{\hbar} + O(B^2),
\end{equation}
with $e$ the electron charge. This allows finally to estimate the diode rectification coefficient $\eta$ in the linear-in-field regime:
\begin{equation}
   \eta =  \frac{\Delta I_{c}}{I_{c}^+ + \left| I_{c}^- \right|} = \frac{2 g^* \mu_B}{\pi \Delta^*} B \equiv \alpha B.
\end{equation}
Using the parameters for InSb ($g^*_{ip} = 25$ and $\Delta^*= 108$~$\mu$eV), we obtain $\alpha = 8.5$~T$^{-1}$ or equivalently a characteristic field $B_0 = 1 / \alpha = 118$~mT.

\begin{figure}[hpt]
    \includegraphics[width=0.7\columnwidth]{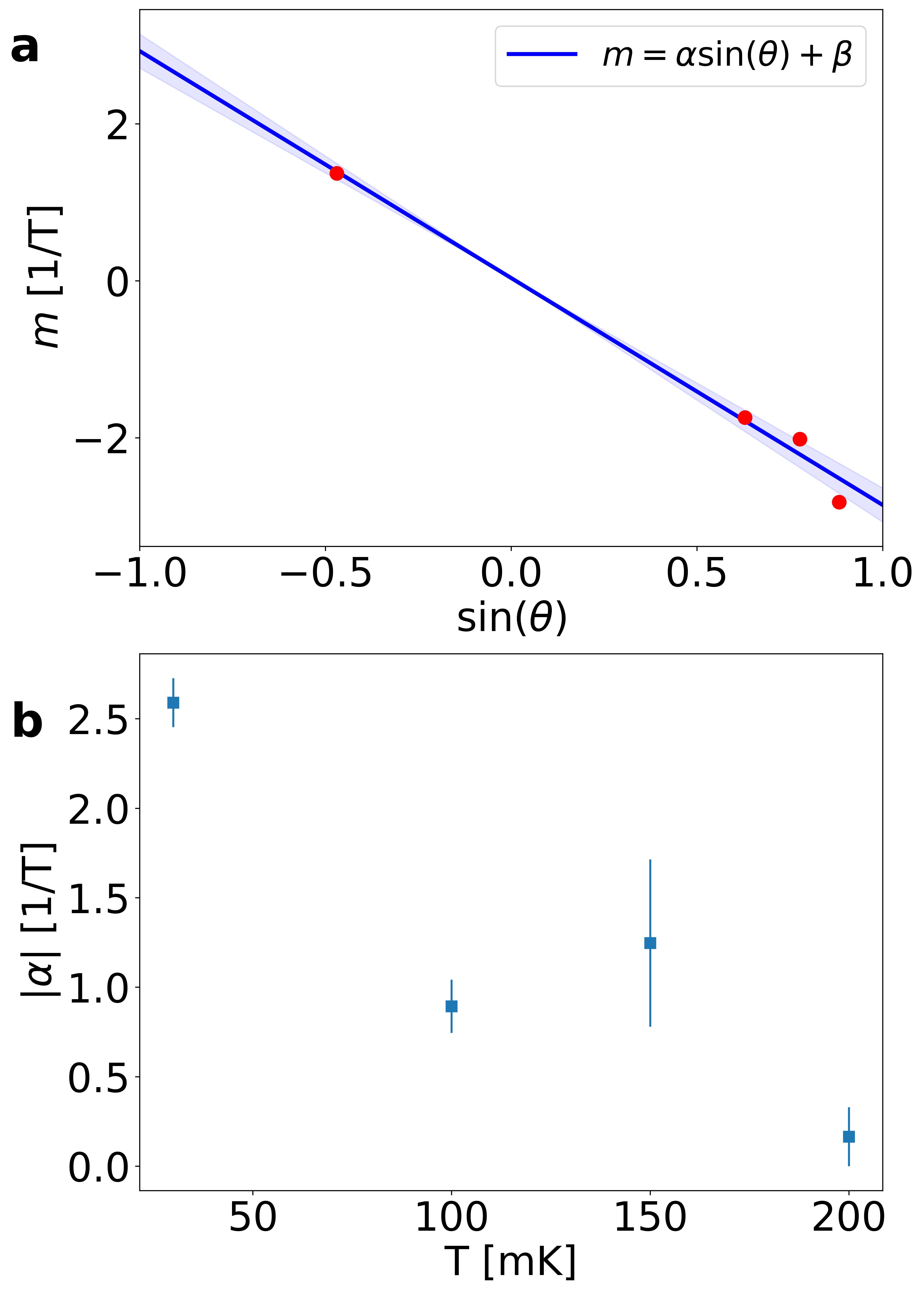}
    \caption{\label{fig:figure5} (a) Proportionality factor $m$ (between rectification coefficient $\eta$ and in--plane magnetic field $B_{ip}$, i.e. $m = \eta / B_{ip}$, see main text), plotted versus the sine of the angle $\theta$ between $B_{ip}$ and the current flow direction. The blue line represents the best linear fit to the data, from which the value of $\alpha$ is extracted. The shaded region indicates the confidence interval. (b) Proportionality factor $\left| \alpha \right|$ plotted versus T, for $\theta = 129$° (Device G5).}
\end{figure}

To compare this result with the experiment, for each curve shown in Figure~\ref{fig:figure4}b, we extract the slope $m$ of the linear fit of $\Delta I_{sw}$ near $B_{ip} = 0$. In Figure~\ref{fig:figure5}a, the values of $m$ are plotted versus the sine of the relative angle $\theta$, normalized to the sum of the two switching currents at zero field (red dots). The blue line is the result of a linear fit, from which the linear coefficient $\alpha = -2.9 \pm 0.2$~T$^{-1}$ is extracted (corresponding to $B_0 = 345$~mT), while the value of the intercept is negligible ($\beta = 0.03$~T$^{-1}$). The relation $m \propto \alpha \sin(\theta)$ indicates that the rectification effect increases with the perpendicular component of the in-plane field $B_{ip,\perp}= B_{ip}\sin(\theta)$.

By considering the behavior of $\Delta I_c$ at finite field predicted in Ref.~\citenum{Davydova2022}, we obtain that the maximum is reached at $q v_F = g^* \mu_B B = \sqrt{16/(\pi^2 + 16)}\; \Delta^* \approx 0.78 \Delta^*$ (see Section X in the SI). Here it would thus be expected to be at $B = 58$~mT, which is higher than experimentally observed. We attribute the discrepancy to the presence of the parallel component of $B_{ip}$, which is expected to suppress the supercurrent flow at higher fields. Moreover, the model does not take into account other effects due to the finite size of the junctions, which could be relevant in our system, as well.

The temperature dependence of the asymmetry curves, shown in Figure~\ref{fig:figure4}e, deserves some attention. In fact, whereas the switching current hardly varies in the temperature range $30-200$~mK, the JDE undergoes a nearly complete suppression. Correspondingly, the rectification coefficient $|\alpha|$ is strongly reduced with increasing temperature (see Figure~\ref{fig:figure5}b). The differing behavior between these two quantities originates from the fact that the JDE is strongly dependent on the presence of higher harmonics in the current-phase relation (CPR) of ballistic SNS junctions \cite{Baumgartner2022}. Indeed, in case of a purely sinusoidal dependence, the anomalous phase shift does not induce any difference between $I_{sw}^+$ and $I_{sw}^-$, which correspond to the maximum and minimum of the CPR, respectively. Higher harmonics decay faster with increasing temperature, so that in the high--temperature limit, the only relevant harmonic is the lowest one, i.e, the CPR is a simple sine function. Thus, the JDE is strongly suppressed in temperature, due to the much stronger dependence of the higher harmonics with respect to the fundamental one.

\subsection{\label{sec:conc}Conclusions}

In conclusion, we have demonstrated that a single planar Josephson junction made from an InSb nanoflag can be driven into the non-reciprocal transport regime by an in--plane magnetic field applied perpendicularly to the direction of the current flow. Moreover, the extent of the rectification depends on the specific combination of the two in-plane field components. Based on symmetry arguments, we have determined that a major role is played by the Rashba-type spin-orbit interaction. Furthermore, we have elucidated the dependence of the effect on other parameters and, specifically, that increasing temperature drastically quenches supercurrent rectification. This is consistent with the absence of higher harmonics in the CPR expected at high temperature.

Thus, high--quality InSb nanoflags are optimal candidates to realize low--dissipation supercurrent rectifiers and to explore the physics of non--reciprocal superconductivity. Further progress in this field will be promoted by the development of microscopic theories which link the rectification quantitatively to the spin--orbit coupling strength. Then, we expect the supercurrent diode effect to become a useful addition to the toolbox of hybrid superconducting electronics.

%%%%%%%%%%%%%%%%%%%%%%%%%%%%%%%%%%%%%%%%%%%%%%%%%%%%%%%%%%%%%%%%%%%%%
%% The "Acknowledgement" section can be given in all manuscript
%% classes.  This should be given within the "acknowledgement"
%% environment, which will make the correct section or running title.
%%%%%%%%%%%%%%%%%%%%%%%%%%%%%%%%%%%%%%%%%%%%%%%%%%%%%%%%%%%%%%%%%%%%%
\begin{acknowledgement}
We thank Daniele Ercolani for his help with the growth of the InSb nanoflags. This research activity was partially supported by the FET-OPEN project AndQC (H2020 Grant No.~828948). A.I., E.S., and F.G.~acknowledge the EU’s Horizon 2020 research and innovation program under Grant Agreement No.~800923 (SUPERTED) and No.~964398 (SUPERGATE) for partial financial support.
\end{acknowledgement}

%%%%%%%%%%%%%%%%%%%%%%%%%%%%%%%%%%%%%%%%%%%%%%%%%%%%%%%%%%%%%%%%%%%%%
%% The same is true for Supporting Information, which should use the
%% suppinfo environment.
%%%%%%%%%%%%%%%%%%%%%%%%%%%%%%%%%%%%%%%%%%%%%%%%%%%%%%%%%%%%%%%%%%%%%
\begin{suppinfo}
Extended data for temperature dependence; Fitting procedure for the dependence $I_{sw}-T$; Experimental data from device G4; Raw data for $\Delta I_{sw}$; JDE in the retrapping current; Parallel component of the in-plane magnetic field; Back-gate dependence of the JDE; Temperature dependence of the JDE; Interference pattern in out-of-plane magnetic field; Universal maximum in $\Delta I_{sw}$; Phenomenological model for $\Delta I_{sw}$ versus $B_{ip}$.
\end{suppinfo}

%%%%%%%%%%%%%%%%%%%%%%%%%%%%%%%%%%%%%%%%%%%%%%%%%%%%%%%%%%%%%%%%%%%%%
%% The appropriate \bibliography command should be placed here.
%% Notice that the class file automatically sets \bibliographystyle
%% and also names the section correctly.
%%%%%%%%%%%%%%%%%%%%%%%%%%%%%%%%%%%%%%%%%%%%%%%%%%%%%%%%%%%%%%%%%%%%%
\bibliography{bianca-sde.bib}

\begin{mcitethebibliography}{68}
\bibitem[foo()]{footnote1}
We note that the data of Ref.~\citenum{Salimian2021}, for which this hysteretic
  behavior was not observed, were obtained at $T = 250$~mK, i.e., at a higher
  temperature than employed here.\relax
\mciteBstWouldAddEndPunctfalse
\mciteSetBstMidEndSepPunct{\mcitedefaultmidpunct}
{}{\mcitedefaultseppunct}\relax
\EndOfBibitem
\end{mcitethebibliography}

\end{document}